\documentclass[aps, pre, amsmath,amssymb]{revtex4}
\usepackage{graphicx}
\usepackage{bm}
\usepackage{epsfig}
\begin{document}
\title{\bf Continuous demixing at liquid-vapor coexistence in a symmetrical 
binary fluid mixture}
\author{Nigel B. Wilding}
\affiliation{Department of Physics, University of Bath, Bath BA2 7AY, United Kingdom}


\begin{abstract} 

We report a Monte Carlo finite-size scaling study of the demixing
transition of a symmetrical Lennard-Jones binary fluid mixture. For
equal concentration of species, and for a choice of the unlike-to-like
interaction ratio $\delta=0.7$, this transition is found to be
continuous at liquid-vapor coexistence. The associated critical end
point exhibits Ising-like universality. These findings confirm those of
earlier smaller scale simulation studies of the same model, but
contradict the findings of recent integral equation and hierarchical
reference theory investigations.


\end{abstract} 
\maketitle
\setcounter{totalnumber}{10}

\section{Introduction and background}

Symmetrical binary fluid mixtures are two-component fluid models in
which the configurational energy is invariant with respect to
interchange of the two particle species. One example is the
symmetrical Lennard-Jones (LJ) mixture in which the interactions
between particles of species 1 and 2 are controlled by a LJ potential
with scale parameters $\sigma_{11}=\sigma_{22} =\sigma_{12}=\sigma$ and
interaction strengths $\epsilon_{11}=\epsilon_{22}=\epsilon\ne
\epsilon_{12}$. The phase diagram of the mixture is uniquely specified
by the ratio of interaction strengths between the unlike and like
species: $\delta \equiv \epsilon_{12}/\epsilon$.

Symmetrical mixtures have been the subject of considerable recent
attention on account of their surprisingly rich phase behaviour
\cite{PAN88,RECHT,DEMIGUEL,WILDING97,WILDING98}. The phase diagram of
the mixture is typically represented in terms of the temperature $T$,
the overall number density $\rho=\rho_1+\rho_2$, and the concentration
$c=\rho_2/(\rho_1 + \rho_2)$. Most attention has focused on the case of
equal species concentration, $c=0.5$. Within this particular symmetry
plane of the full phase diagram, the $\rho$-$T$ dependence of the phase
behaviour on the value of $\delta$ has been systematically studied
using simulation and mean field theory for the case of a symmetrical
square-well mixture \cite{WILDING98}. The findings of that work are
summarized in fig 1. of ref.~\cite{WILDING98}. Three
distinct classes of phase diagram were found depending on the choice of
the parameter $\delta$. Specifically, for large $\delta<1$
(fig. 1(a) of ref.~\cite{WILDING98}) there occurs a
``$\lambda$-line'' of critical demixing transitions which intersects
(and is truncated by) the liquid-vapor line at a critical end point
(CEP). For small $\delta>0$, on the other hand, the line of demixing
transitions intersects the liquid-vapor line at the liquid-vapor
critical critical point, forming a tricritical point
(fig. 1(c) of ref.~\cite{WILDING98}). Intermediate between these two regimes, one
observes both a liquid-vapor critical point and a tricritical point at
higher densities (fig. 1(b) of ref.~\cite{WILDING98}). In the simulations this
latter regime was observed to occur for $0.65\lesssim \delta \lesssim
0.68$, while within the particular mean field theory employed it
occurred for $0.605\lesssim \delta \lesssim 0.708$.

It has been suggested \cite{WILDING98} that the scenario of phase
behaviour outlined above for the square well mixture, might in fact be
generic to all symmetrical fluids. On the theoretical side, support for
this proposal has come from studies of variety of model systems,
including mean field studies of a lattice fluid model \cite{SCHOEN} and
integral equation theories
\cite{KAHL01,PASCHINGER1,PASCHINGER2,CACCAMO} of the hard core Yukawa
fluid mixture (HCYFM). The same overall scenario was initially reported
for the hard core $+$ Lennard-Jones mixture by Pini {\em et al}
\cite{PINI00} in a study employing the Hierarchical Reference Theory
(HRT)-- a powerful fluctuation-inclusive approach based on
renormalization group techniques \cite{PAROLA91}. By contrast, the
available simulation data for models other than the square well fluid
is less comprehensive and has, in the main part, concentrated on the
regime of large $\delta$. Specifically, MC simulations of the
symmetrical LJ fluid \cite{WILDING97} found evidence for a CEP (and
associated coexistence curve singularities) for the particular case
$\delta=0.7$. Similarly a CEP has been reported in simulation studies
of the HCYFM for $\delta=0.9$ \cite{CACCAMO}.

Very recently, however, the situation with specific regard to the CEP
regime has become less clear following two theoretical studies which
question its existence in the expected range of $\delta$. In the first
of these, Antonevych {\em et al.} \cite{ANTON02} applied the modified
hypernetted chain (MHNC) integral equation theory to the symmetrical LJ
fluid, but found no evidence for a CEP at $c=0.5$. Instead they
reported that the intermediate regime  (fig. 1(b) of
ref.~\cite{WILDING98}) persists right up to $\delta=0.81$, (the limit
of their study). This value of $\delta$ is considerably greater than
that ($\delta=0.7$) at which CEP behaviour was originally found in the
simulations of the same model. On this basis, Antonevych {\em et al.}
suggested that the demixing transition could always be first order at
liquid-vapor coexistence, although they speculated that it might be
only weakly so. Furthermore, they questioned the ability of simulation
to distinguish this possibility from a true CEP.

In a separate study, Pini {\em et al.} \cite{PINI02} have applied the
HRT approach to the HCYFM. They too observed no CEP,  instead reporting
(in accord with the finding of Antonevych {\em et al.} \cite{ANTON02})
that the intermediate regime persists (albeit weakly) to $\delta=0.8$,
the limit of their study. Moreover, these authors report \cite{NOTE} that in
their previous study of the LJ mixture \cite{PINI00}, the apparent
observation of a CEP was an artifact stemming from the low
resolution of the grid in the $\rho$-$c$ plane on which the HRT
equations were solved.

Clearly, therefore, the evidence emerging from the various theoretical
and simulation studies is contradictory with regard to the existence of
a CEP and the range of values of $\delta$ for which it occurs. Most
significant in our view is the failure of HRT to observe a CEP for
$\delta <0.8$. Given the fluctuation-inclusive nature of this approach
(and its success in other contexts, see e.g.  \cite{WILDING01}), one
cannot simply ascribe discrepancies between the simulations and the
MHNC study of the LJ fluid \cite{ANTON02} to the neglect of critical
fluctuations by the latter.

Questions have been raised too regarding the reliability of simulation
in distinguishing a CEP from a weak first order transition
\cite{ANTON02}. While, in our opinion, the original
work~\cite{WILDING97} {\em does} provide compelling evidence for CEP
behaviour, it is certainly the case that owing to the prevailing
computational constraints, the range of system sizes studied was
smaller than one might have wished. This in turn precluded an investigation of
corrections to scaling and the approach to the limiting critical
behaviour.

In the light of the above considerations, it would seem worthwhile to
revisit the original system in which a CEP was observed, with a view to
performing a more comprehensive determination of the nature of the
demixing transition at liquid-vapor coexistence. To this end we have
carried out a detailed Monte Carlo (MC) finite-size scaling (FSS) 
simulation study of the symmetrical LJ fluid with $\delta=0.7$, for
which it was previously claimed that a CEP occurs \cite{WILDING97}. The
new study has been executed along similar lines to the original one,
but employed a considerably larger range of systems sizes, thus
permitting a more sophisticated finite-size scaling analysis.  The
results demonstrate (unambiguously, we believe) that the model does
indeed exhibit a CEP of the expected Ising type.

\section{Method and results}

The simulation methodology employed here is broadly similar to that
described in ref.~\cite{WILDING97} and we refer the reader to that
paper for a detailed description. Briefly, we have performed a MC
simulation study of the liquid-vapor coexistence line of the
symmetrical LJ fluid mixture with $\delta=0.7$ in the neighborhood of
the demixing transition. The simulations were carried out within the
grand canonical ensemble and the chemical potentials of the two species
were constrained to be equal ($\mu_1=\mu_2=\mu$), implying on symmetry
grounds that $\langle c\rangle=0.5$. The system was confined to a cubic
box of linear dimension $L$ having periodic boundary conditions. In
all, seven system sizes were studied having $L=10\sigma, 12.5\sigma,
15\sigma, 17.5\sigma, 20\sigma, 25\sigma$ and $30\sigma$. The largest
of these systems contains some $1.6\times 10^4$ particles in the liquid
phase at the demixing point. The range of system sizes studied here is
to be compared with that of the original study \cite{WILDING97}, for
which the largest size attained was $L=12.5\sigma$. The greater range
of sizes studied here was partly facilitated by the use of high
performance parallel computers.

The liquid-vapor coexistence curve $\mu_{\rm cx}(T)$ was obtained using
multicanonical extended sampling techniques \cite{BERG}. These allow
both the liquid and vapor phases to be sampled (and thus connected) in
a single simulation run. A suitable form for the requisite preweighting
function was obtained using a variant of the recently proposed
Wang-Landau method \cite{WANGLAN}. To locate the coexistence line, the
equal peak area criterion was applied to the measured form of the
number density distribution function $p_L(\rho)$ \cite{EW,WILDING95}.
Histogram reweighting techniques were employed to fine-tune this
procedure and to facilitate the mapping of a portion of the
liquid-vapor coexistence curve in the neighborhood of the demixing
transition.

\begin{figure}
\includegraphics[width=10.0cm,clip=true]{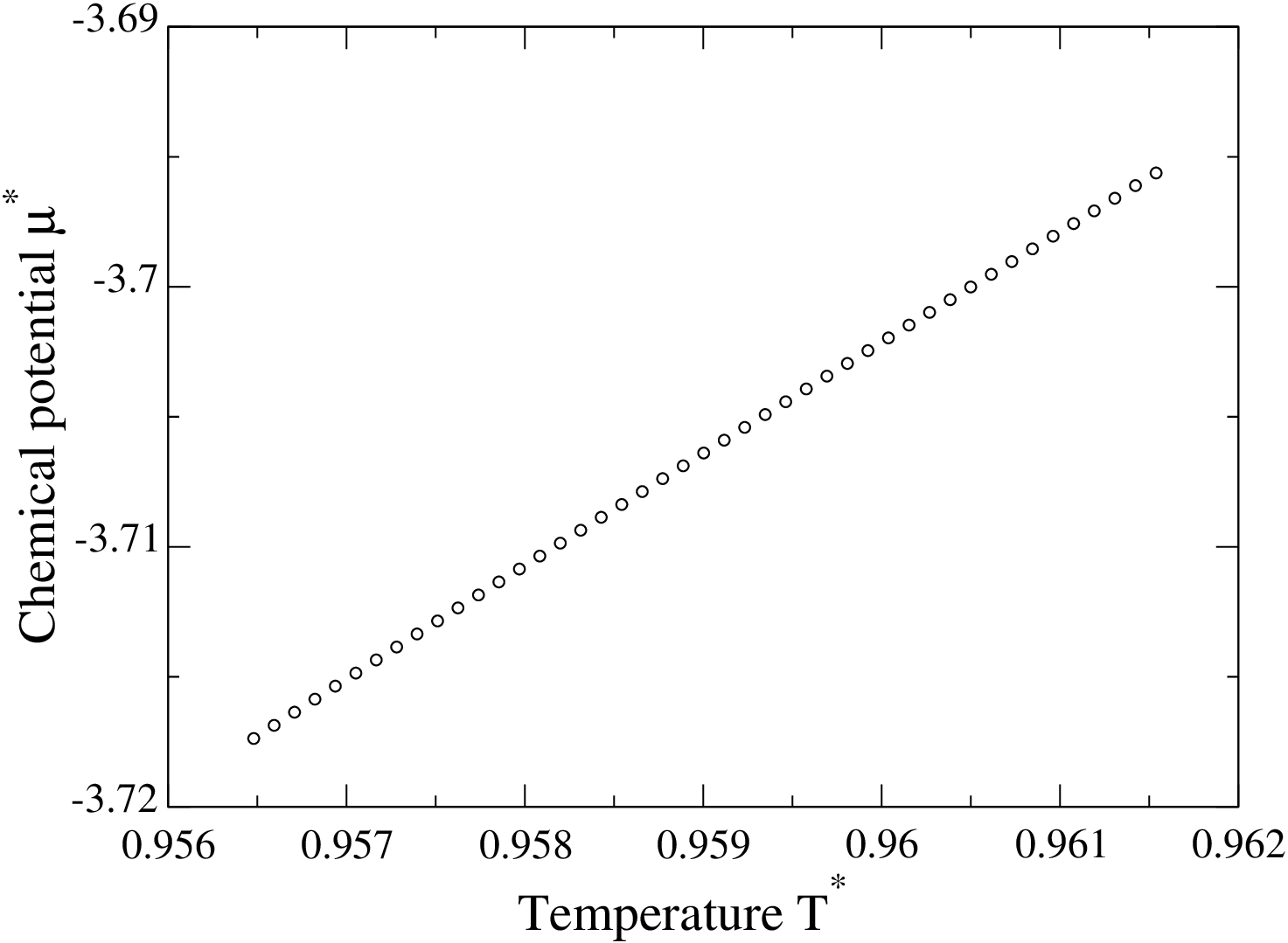}

\caption{The liquid-vapor coexistence curve in the neighborhood of
the demixing transition, obtained for the $L=25\sigma$ system. Statistical
errors are considerably smaller than the symbol sizes.}

\label{fig:cxcurve}
\end{figure}

\begin{figure}
\includegraphics[width=10.0cm,clip=true]{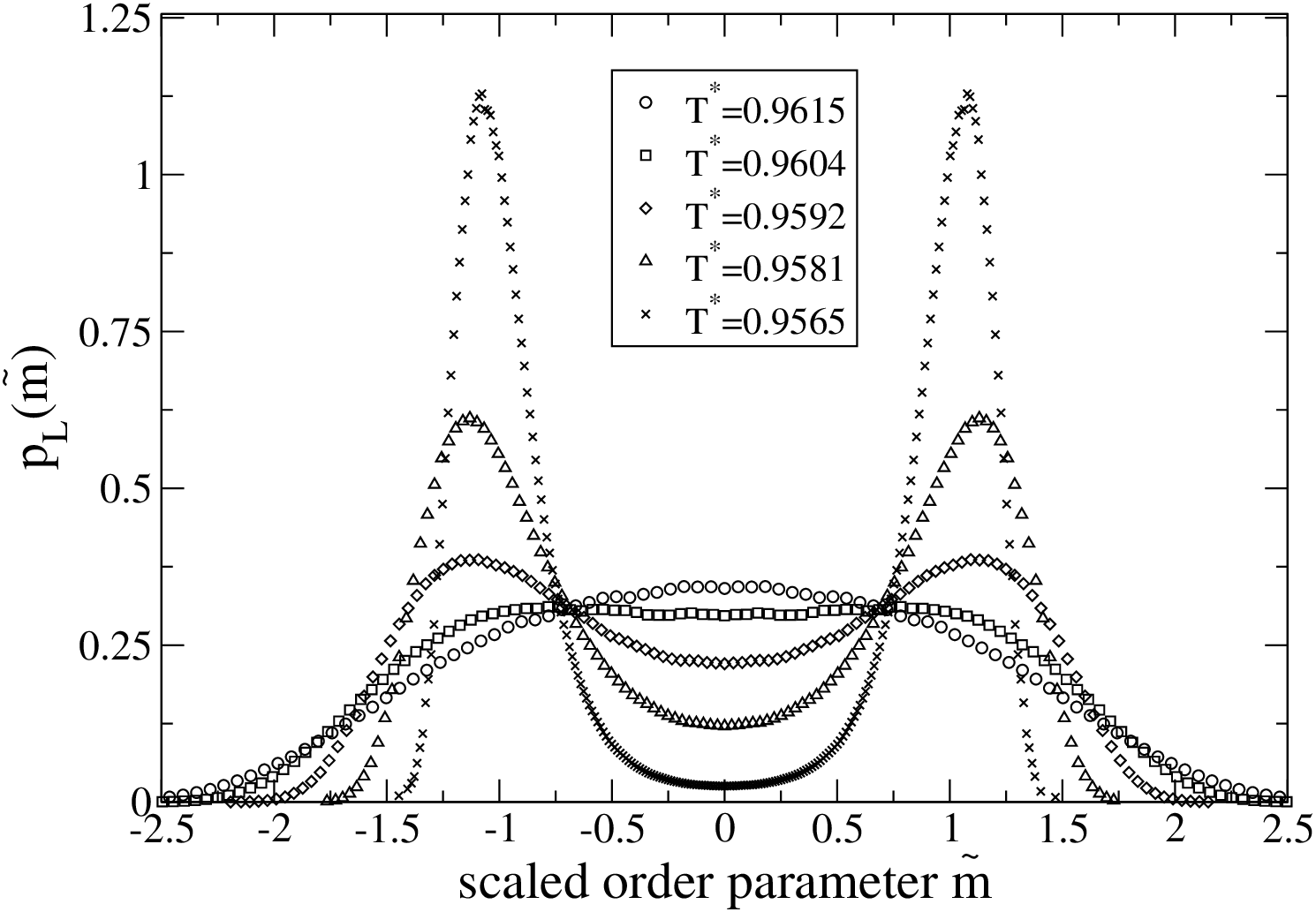}

\caption{The distribution of the order parameter obtained for the
$L=25\sigma$ system at a number of temperatures along the liquid branch
of the coexistence curve. In each case, the order parameter has been
scaled $m\to\tilde{m}$ to ensure that all distributions have unit
variance.}

\label{fig:concdist}
\end{figure}

Fig.~\ref{fig:cxcurve} shows the measured form of the liquid-vapor
coexistence line $\mu_{\rm cx}(T)$ \cite{NOTE0}. To elucidate the
nature of the demixing transition along the liquid branch of this
tightly determined coexistence line, we have measured the probability
distribution function $p_L(m)$ of the demixing order parameter $m$,
where $m=(N_1-N_2)/(N_1+N_2)$, with $N_1$ and $N_2$ the
instantaneous counts of the respective particle species. The form of
$p_L(m)$, for the $L=25\sigma$ system at temperatures spanning the
demixing transition, is shown in fig.~\ref{fig:concdist}. We find that
$p_L(m)$ evolves smoothly between the strongly double peaked form
(corresponding to a demixed liquid) at low temperature and a singly
peaked form (corresponding to a mixed liquid) at higher temperatures.

\begin{figure}
\includegraphics[width=10.0cm,clip=true]{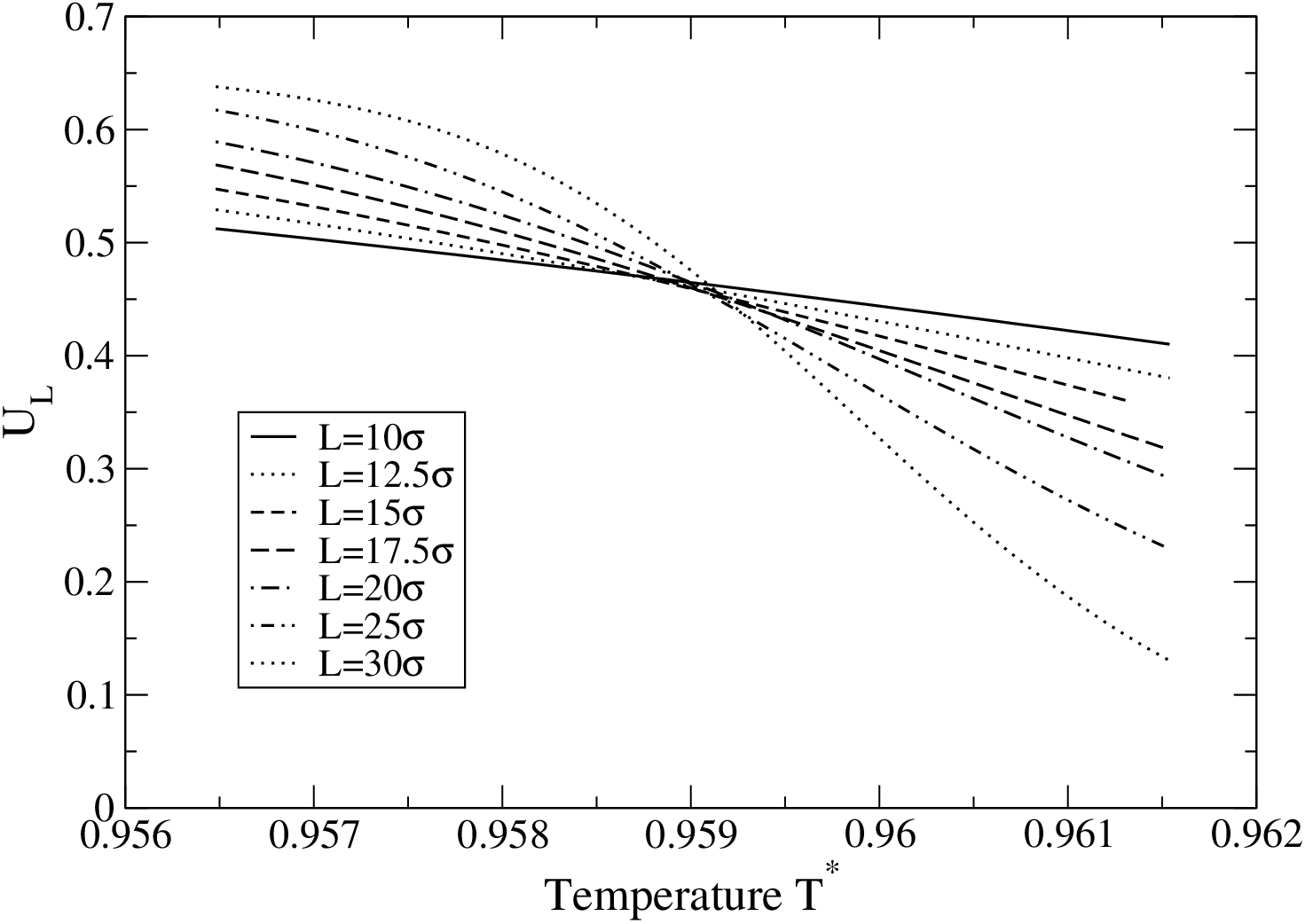}

\caption{The temperature dependence of the fourth order cumulant ratio (see
text) measured along the liquid branch of the coexistence curve.}

\label{fig:cums}
\end{figure}

To quantify this evolution more precisely, we have measured the
temperature dependence of the fourth order cumulant ratio:
$U_L=1-\langle m^4\rangle/3\langle m^2\rangle^2$, which provides a
dimensionless measure of the shape of a distribution. The results
(fig~\ref{fig:cums}) demonstrate (for all system sizes) a smooth evolution of
the concentration distribution from the ordered to the disordered
phase. The rate of change increases with the system size, but there is
no evidence of a jump discontinuity. More significantly, there is a
fairly well defined crossing point of the curves for $U_L\approx 0.46$
close to the reduced temperature $T^\star\equiv 1/\epsilon=0.959$. Such
an intersection reflects scale invariance in the concentration
distribution and does not occur at a first order phase transition. We
note further that the magnitude of the cumulant ratio at the
intersection point corresponds closely to the known Ising universal
value, $U_L^{\rm Ising}=0.465$ \cite{TSYPIN00}.

\begin{figure}
\includegraphics[width=10.0cm,clip=true]{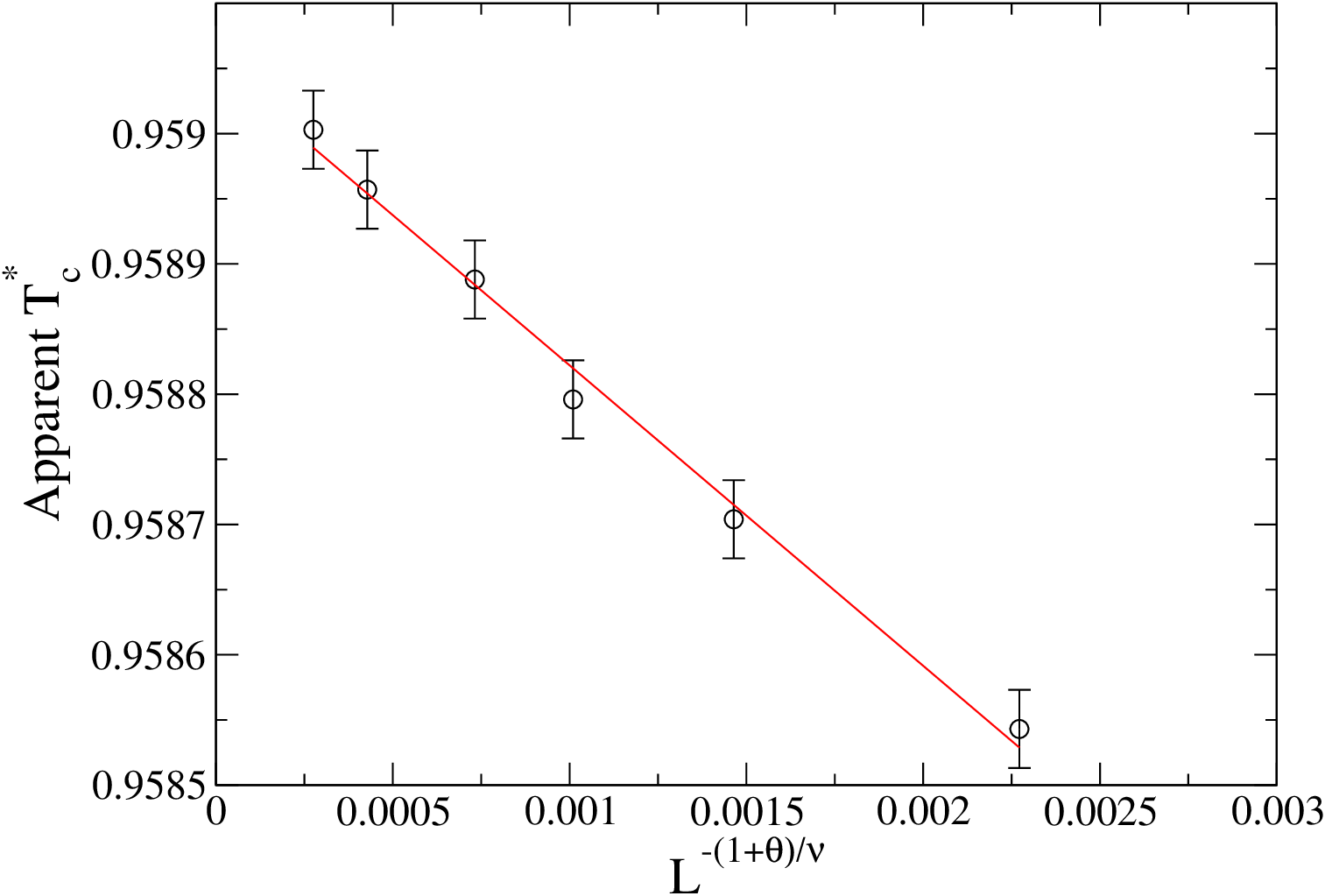}

\caption{The finite-size scaling behaviour of the corrections to scaling
obtained according to the method described in the text and
ref.~\protect\cite{WILDING95}. The straight line is a linear fit to the data.
}

\label{fig:tc_scale}
\end{figure}

Although the results of fig.~\ref{fig:cums} constitute strong evidence
for a critical demixing transition, a close inspection reveals that the
cumulant crossings do not occur precisely at a unique temperature for
all system sizes. Such behaviour mirrors that observed in other
contexts (see eg. \cite{WILDING95,WILDING01}) where it was found to be
attributable to corrections to finite-size scaling. Since these
corrections have a known universal scaling form,  one can attempt to
fit them, thereby permitting an extrapolation to the thermodynamic
limit. The procedure involves determining, for each system size, the
coexistence state point for which the form of $p_L(m)$ on the liquid
branch best matches the (independently known \cite{TSYPIN00}) fixed
point order parameter distribution function $p^*(m)$ appropriate to the
Ising universality class. The matching temperature as a function of
system size is expected to scale like $T_c(L)\sim L^{-(1+\theta)/\nu}$
\cite{WILDING95}, where $\theta\approx 0.50(2)$ is the Ising
correction-to-scaling exponent \cite{JUSTIN} and $\nu=0.6294(4)$ is the
correlation length exponent \cite{FERRENBERG}. Fig.~\ref{fig:tc_scale}
shows the data expressed in this form, which indeed exhibits the
anticipated scaling behaviour. Extrapolation of a linear fit to
$L=\infty$ yields an estimate for the critical temperature $T_{\rm
cep}=0.9595(3)$, which occurs for $\mu_{\rm cep}=-3.704(2)$.

\begin{figure}
\includegraphics[width=10.0cm,clip=true]{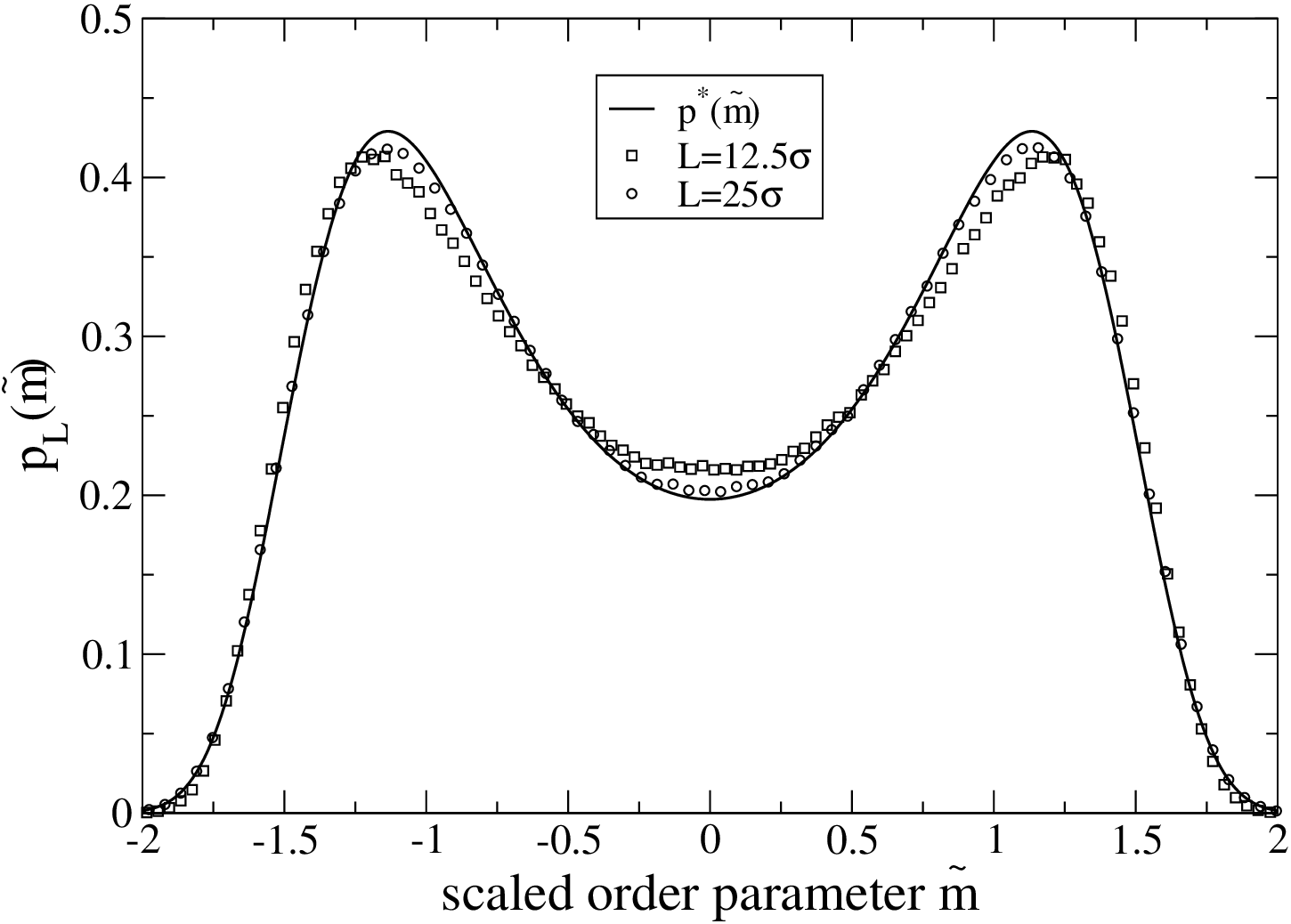}

\caption{Order parameter distribution for the $L=12.5\sigma$ and
$L=25\sigma$ system sizes at the extrapolated parameters of the critical end
point (cf. fig.\protect\ref{fig:tc_scale}). Also shown (solid line) is
the limiting universal fixed point form $p^\star(\tilde {m})$
\protect\cite{TSYPIN00}.}

\label{fig:dists_tc}
\end{figure}

It is instructive to plot the finite-size forms of $p_L(m)$ at the
estimated critical point in order to expose the scale and character of
corrections to finite-size scaling. Figure~\ref{fig:dists_tc} compares
the distributions for the $L=12.5\sigma$ and $L=25\sigma$ system sizes
with that of $p^\star (m)$ \cite{TSYPIN00}. One sees that the results
for the larger system size closely match the limiting form.

\begin{figure}
\includegraphics[width=10.0cm,clip=true]{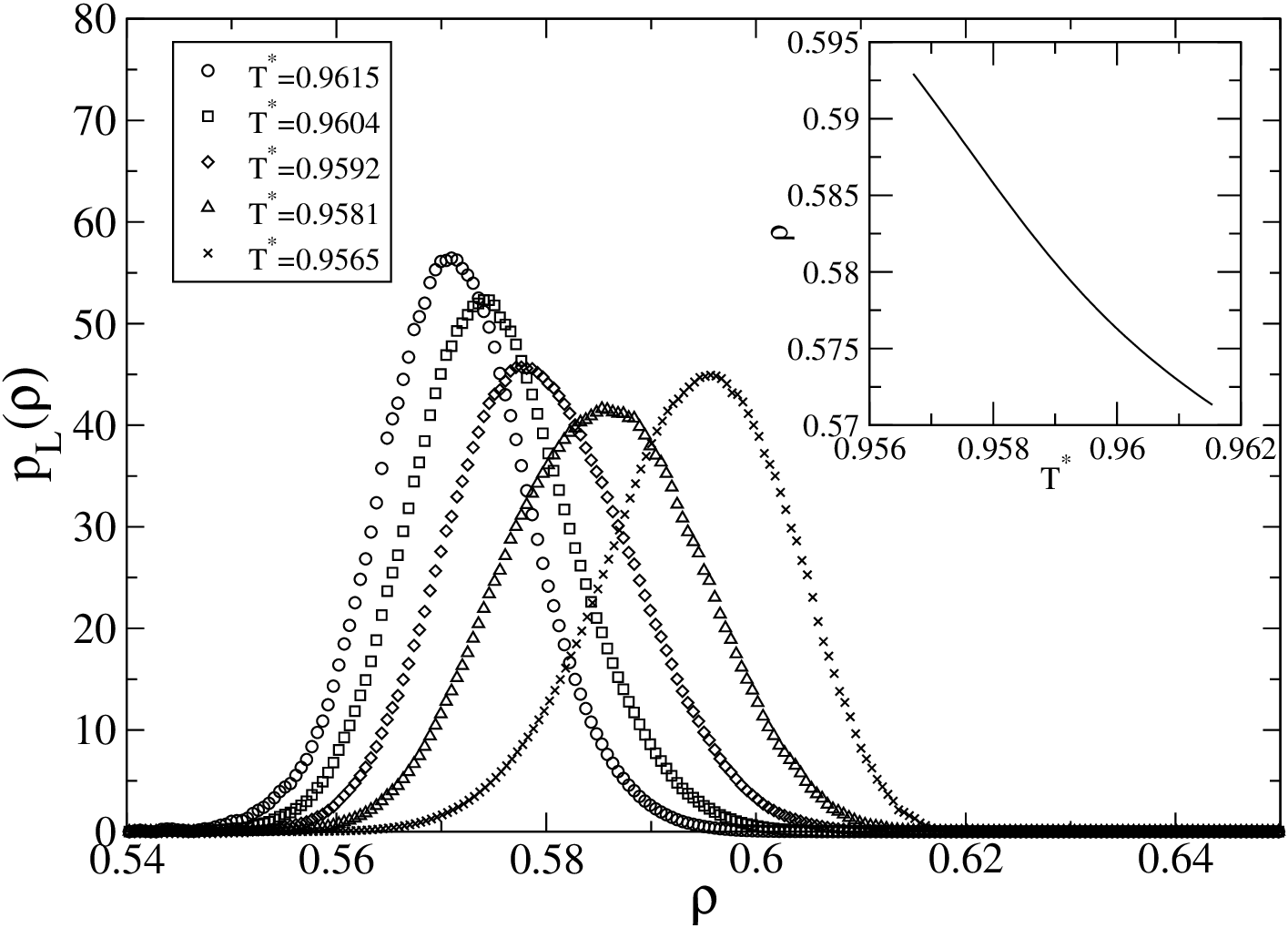}

\caption{The measured form of $p_L(\rho)$ on the liquid branch for the
$L=25\sigma$ system size for a number of temperature spanning the
demixing point, The inset shows the corresponding average number
density as a function of temperature.}

\label{fig:dendists}
\end{figure}

Finally, we have considered the behaviour of the number density along
the liquid branch at coexistence. Were the demixing transition first
order along the coexistence line, one would expect that coupling
between the concentration and the number density would engender a first
order phase transition in the liquid branch number density
(cf. fig. 1(b) of ref.~\cite{WILDING98}). This would be manifest as a 
discontinuity in the density of the coexisting liquid phase and by a
splitting of the liquid-phase peak in the density distribution
$p_L(\rho)$. Fig.~\ref{fig:dendists} shows the measured form of the
liquid-phase peak for the $L=25\sigma$ system size, for several
coexistence state points spanning the demixing temperature. Also shown
(inset) is the average liquid phase density. The distribution of the
density in the liquid phase is singly peaked and we find no evidence of
a discontinuity in the average density for any of the system sizes
studied \cite{NOTE1}.

\section{Discussion and conclusions}

In summary, we have performed a finite-size scaling analysis of the
demixing transition of a symmetrical LJ fluid along the liquid branch
of the liquid-vapor coexistence line. The results demonstrate that for
the case $\delta=0.7$, this transition is continuous and of the
expected Ising type \cite{DIEHL}. Accordingly we can conclude that this
system does indeed exhibit a critical end point, as originally proposed
in ref~\cite{WILDING97}. This clear finding should provide a benchmark
for testing the reliability of theoretical approaches to the phase
behaviour of symmetrical mixtures.

With regard to the original motivation for this study, namely the
failure of both the MHNC integral equation study and HRT to observe a
CEP in the expected range of $\delta$ \cite{ANTON02,PINI02}, it is
difficult to comment definitively at this stage on the source of the
discrepancies. We note, however, that problems with the accuracy of the
MHNC approach for studying the liquid branch of the HCYFM have
previously been reported \cite{CACCAMO}. With regard to the HRT, it
would certainly be useful if this method could be applied to other
symmetrical fluid models in order to determine whether there is any
systematic model dependence on the qualitative features of the phase
behaviour.



\newpage





\begin{thebibliography}{99}
\bibitem{NOTE0} Owing to a conversion error, the values of the
coexistence chemical potentials quoted in \protect\cite{WILDING97} were
displaced from their true value by $\ln 2$.
Figure~\protect\ref{fig:cxcurve} corrects that error.
\end{thebibliography}
\end{document}